\documentclass[aps,prev,preprintnumbers,floatfix,nofootinbib]{revtex4-1}
\pdfoutput=1

\usepackage{amsmath}
\usepackage{amsfonts}
\usepackage{amssymb}
\usepackage{graphicx}
\usepackage{bm}
\usepackage{times}
\usepackage{hyperref}
\usepackage{slashed}
\usepackage{color}

\begin{document}

\title{Brazilian Community Report on Dark Matter}

\author{E. Abdalla$^1$}
\author{I. F. M. Albuquerque$^1$}
\author{A. Alves$^2$}
\author{L. Barosi$^3$}
\author{M. C. Q. Bazetto$^4$}
\author{R. C. Batista$^5$}
\author{C. A. Bernardes$^6$}
\author{C. Bonifazi$^7$}
\author{H. A. Borges$^8$}
\author{F. A. Brito$^{3,9}$}
\author{T. R. P. Caram\^es$^{10}$}
\author{L. Casarini$^{11}$}
\author{D. Cogollo$^3$}
\author{A. G. Dias$^{12}$}
\author{A. Esmaili$^{13}$}
\author{M. M. Ferreira$^{14}$}
\author{G. Gil da Silveira$^{15,16}$}
\author{M. M. Guzzo$^{17}$}
\author{D. Hadjimichef$^{16}$}
\author{P. C. de Holanda$^{17}$} 
\author{E. Kemp$^{17}$}
\author{A. Lessa$^{12}$}
\author{G. Lichtenstein$^1$}
\author{A. A. Machado$^{17}$}
\author{M. Makler$^{18}$}
\author{V. Marra$^{19,20}$}
\author{R. D. Matheus$^{21}$}
\author{P. G. Mercadante$^{12}$}
\author{T. B. de Melo$^{3,22}$}
\author{C. Nishi$^{12}$}
\author{A. Nepomuceno$^{23}$}
\author{S. F. Novaes$^{21}$}
\author{V. L. Pimentel$^{5,24}$}
\author{P. R. D. Pinheiro$^{14}$}
\author{C. A. Pires$^{3}$}
\author{A. R. Queiroz$^{3}$}
\author{F. S. Queiroz\footnote{Contact Editor: farinaldo.silva.queiroz@gmail.com}$^{22}$}
\author{M.~S.~Rangel$^7$}
\author{D. C. Rodrigues$^{19,20}$}
\author{J. G. Rodrigues$^{25}$}
\author{V. de Souza$^{26}$}
\author{P. S. Rodrigues da Silva$^{3}$}
\author{C. Siqueira$^{22}$}
\author{E. Segreto$^{17}$}
\author{B. L. Sanchez-Vega$^{27}$}
\author{R. Rosenfeld$^{21}$}
\author{J. R. L. Santos$^{3}$}
\author{A.C.O. Santos$^{22}$}
\author{R. Silva$^{25,28}$}
\author{D. Sokolowska$^{22}$}
\author{P. R. Teles$^{29}$}
\author{T. R. F. P. Tomei$^{21}$}
\author{G. A. Valdiviesso$^{30}$}
\author{P. Vasconcelos$^{3}$}
\author{A. Viana$^{26}$}

\affiliation{$^1$Instituto de F\'isica, Universidade de S\~ao Paulo, Brazil}


\affiliation{$^2$Departamento de F\'isica, Universidade Federal de S\~ao Paulo, UNIFESP, Diadema, 09972-270, Brazil}

\affiliation{$^3$Departamento de F\'isica, Universidade Federal de Campina Grande,
Campina Grande, PB, Brazil}

\affiliation{$^4$Centro de Tecnologia da Informação-CTI Renato Archer, Rd. D.Pedro, Km 143,6, Campinas, SP. CEP 13069-901}


\affiliation{$^5$Escola de Ci\^encias e Tecnologia, Universidade Federal do Rio Grande do Norte Campus Universit\'ario Lagoa Nova, Natal, RN, Brazil, CEP 59078-970}


\affiliation{$^6$N\'ucleo de Computação Científica, Universidade Estadual Paulista, Rua Dr. Bento Teobaldo Ferraz, 271, CEP 01140-070 S\~{a}o Paulo, SP,
Brazil
} 

\affiliation{$^7$Instituto de F\'isica, Universidade Federal do Rio de Janeiro (UFRJ),Caixa Postal 68528, CEP 21941-972, Rio de Janeiro, RJ, Brazil}

\affiliation{$^8$Instituto de F\'isica, Universidade Federal da Bahia, 40210-340, Salvador, BA, Brasil}

\affiliation{$^{9}$Departamento de F\'isica, Universidade Federal da Para\'iba, Caixa Postal 5008, 58051-970, Jo\~ao Pessoa, PB, Brazil}


\affiliation{$^{10}$Departamento de F\'isica, Universidade Federal de Lavras, Caixa Postal 3037, 37200-900 Lavras, MG, Brazil}

\affiliation{$^{11}$Departamento de F\'isica, Universidade Federal de Sergipe, 49100-000, Aracaju, SE, Brasil}

\affiliation{$^{12}$Centro de Ci\^encias Naturais e Humanas, Universidade Federal do ABC, 09210-580, Santo Andr\'e-SP, Brazil}


\affiliation{$^{13}$Departamento de F\'isica, Pontif\'icia Universidade Cat\'olica do Rio de Janeiro, Rio de Janeiro 22452-970, Brazil}

\affiliation{$^{14}$Departamento de F\'isica, Universidade Federal do Maranh\~ao, Campus Universit\'ario do Bacanga, S\~ao Luis - MA, 65080-805 - Brazil}

\affiliation{$^{15}$Instituto de F\'{\i}sica, Universidade Federal do Rio Grande do
  Sul, Av. Bento Gon\c{c}alves, 9500\\
Porto Alegre, Rio Grande do Sul, CEP 91501-970,
Brazil}


\affiliation{$^{16}$Departamento de F\'{\i}sica Nuclear e de Altas Energias, Universidade do Estado do Rio de Janeiro\\
CEP 20550-013, Rio de Janeiro, RJ, Brazil}

\affiliation{$^17$Instituto de F\'sica Gleb Wataghin - UNICAMP, 13083-859, Campinas SP, Brazil}

\affiliation{$^{18}$Centro Brasileiro de Pesquisas F\'isicas, Rio de Janeiro, RJ, Brazil}

\affiliation{$^{19}$Center for Astrophysics and Cosmology, CCE, Federal University of Esp\'irito Santo, 29075-910, Vit\'oria, ES, Brazil}

\affiliation{$^{20}$Department of Physics, CCE, Federal University of Esp\'irito Santo, 29075-910, Vit\'oria, ES, Brazil}

\affiliation{$^{21}$Instituto de F\'isica Te\'orica, Universidade Estadual Paulista, SP, Brazil}

\affiliation{$^{22}$International Institute of Physics,  Universidade Federal do Rio Grande do Norte,Campus  Universit\'ario,  Lagoa  Nova,  Natal-RN  59078-970,  Brazil}

\affiliation{$^{23}$Departamento de Ci\^encias da Natureza Universidade Federal Fluminense Rua Recife s/n,Rio das Ostras,
Rio de Janeiro, Brazil}


\affiliation{$^{24}$Faculdade de Engenharia El\'etrica e de Computa\c c\~ao-FEEC. Av. Albert Einstein, 400 - Bar\~ao Geraldo, Campinas - SP, 13083-852}


\affiliation{$^{25}$Universidade Federal do Rio Grande do Norte, Departamento de F\'isica, Natal - RN, 59072-970, Brazil}

\affiliation{$^{26}$Instituto de F\'{\i}sica de S\~ao Carlos, Universidade de S\~ao Paulo.}

\affiliation{$^{27}$Universidade  Federal  de  Minas  Gerais,  ICEx,  Dep.   de  F\'isica ,Av. Ant\^ onio  Carlos,  6627,  Belo  Horizonte,  MG,  Brasil,  CEP  31270-901}

\affiliation{$^{28}$Departamento de F\'isica, Universidade do Estado do Rio Grande do Norte, Mossor\'o, 59610-210, Brasil}

\affiliation{$^{29}$Departamento de F\'isica Te\'orica - IF - UERJ - Rua S\~ao Francisco Xavier 524, Rio de
Janeiro, RJ, Maracan\~a, CEP:20550}

\affiliation{$^{30}$Instituto de Ci\^encia e Tecnologia, Universidade Federal de Alfenas, Campus Po\c cos de Caldas, 37715-400, Po\c cos de Caldas - MG, Brazil}

\begin{abstract}
\vspace{5cm}
This white paper summarizes the activities of the Brazilian community concerning dark matter physics and highlights the importance of financial support to Brazilian groups that are deeply involved in experimental endeavours. The flagships of the Brazilian dark matter program are the Cherenkov Telescope Array, DARKSIDE, SBN and LHC experiments, but we emphasize that smaller experiments such as DAMIC and CONNIE constitute important probes to dark sectors as well and should receive special attention. Small experimental projects showing the potential to probe new regions of parameter space of dark matter models are encouraged. On the theoretical and phenomenological side, some groups are devoted to astrophysical aspects such as the dark matter density profile while others explore the signature of dark matter models at colliders, direct and indirect detection experiments. In summary, the Brazilian dark matter community that was born not long ago has grown tremendously in the past years and now plays an important role in the hunt for a dark matter particle.
\end{abstract}

\maketitle
\flushbottom

\section{Introduction}

The evidence for dark matter stems from different times and distance scales of our Universe.  The precise measurements of the cosmic microwave background (CMB) power spectrum, galaxy rotation curves, galaxy clusters, gravitational lensing, baryonic acoustic oscillations, big bang nucleosynthesis have shaped our understanding of the universe and represent a whole panoply of evidence for dark matter.  However, all these data are gravitational informing us not much about the nature of dark matter.  Consequently, dark matter could be comprised of particles whose masses span over orders of magnitude from $10^{-22}$~eV to $10^{12}$~eV (1~PeV) (See Fig.~\ref{fig:proudBrazilians}).

Despite our ignorance concerning the mass of the dark matter particle we have accumulated important information concerning dark matter, such that any dark matter candidate should fulfill some requirements: (i) should yield the correct relic density; (ii) should be non-relativistic at matter-radiation equality to form structures in the early Universe in agreement with the observation; (iii) it should be effectively neutral otherwise it would form  unobserved  stable charged particles; (iv) and last but not least, it should be cosmologically stable with a lifetime much larger than the age of the universe to be consistent with cosmic rays and gamma-rays observations. Having these requirements in the back of our minds three classes of candidates naturally arise: 

\begin{figure}[h!]
\centering
\includegraphics[scale=0.3]{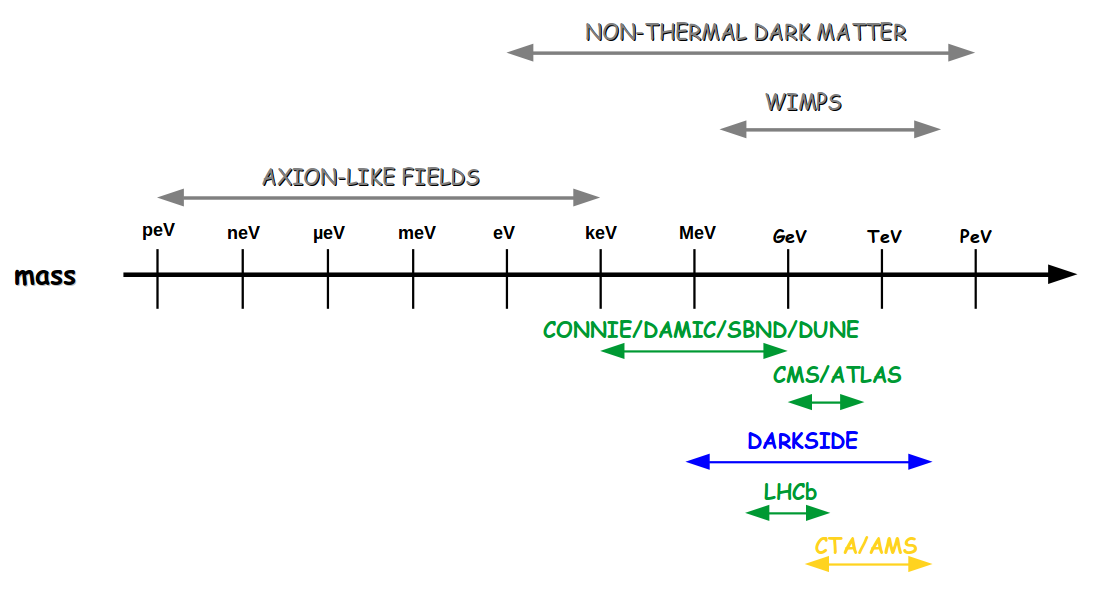}
\caption{Mass range of the dark matter candidates and the corresponding experiments with Brazilian involvement.}
\label{fig:proudBrazilians}
\end{figure}

\begin{itemize}
    \item {\bf WIMPs}
    
WIMPs (Weakly Interacting Massive Particles) appear in several popular model building setups. They are based on the thermal decoupling paradigm which is a key input in the observables such as the Cosmic Microwave Background and Big Bang Nucleosynthesis.  Thus, a first guess is to assume that dark matter particles also belong to a thermal history. Under these assumptions, the correct relic density is obtained for an annihilation cross-section of the order of the electroweak scale. That said, assuming a standard thermal history in the universe and that the overall dark matter relic density is governed solely by the dark matter annihilation cross-section into Standard Model particles we conclude that the dark matter is restricted  to be within a narrow mass range as shown in Fig.\ref{fig:proudBrazilians}. Typically the parameters that govern the relic density dictate the direct and indirect detection signatures as well resulting in an exciting dark matter complementary program \cite{Akerib:2015rjg,Ahnen:2016qkx,Abdallah:2016ygi,Amole:2017dex,Aprile:2018dbl,Ajaj:2019imk,Aprile:2019dbj} . There are ways to break this complementary study while reproducing the correct dark matter relic density but we will not address them here. Anyway, usually WIMP models that reproduce the correct relic density also produce signals that are within reach of current and upcoming experiments \cite{Arcadi:2017kky}. As WIMPs arise naturally in several model building endeavours and are entitled to a rich phenomenology they continue to be quite appealing despite the null results so far.  

    \item {\bf Axions}
    
Axions are also very popular dark matter candidates for being related to a peculiar solution of the strong CP problem \cite{Peccei:1988ci,Kim:2008hd}. The non-perturbative dynamics of the strong interaction sector of the Standard Model leads to an effective Lagrangian ${\mathcal{L}}_{eff}=\frac{\bar{\theta}}{32\pi^{2}}\epsilon_{\mu\nu\sigma\rho}G_{a}^{\mu\nu}G_{a}^{\sigma\rho}$, in which $G_{a}^{\mu\nu}$ is gluon field strength tensor, $\epsilon_{\mu\nu\sigma\rho}$ the totally anti-symmetric tensor and  $\bar{\theta}$ is a free parameter, which also includes a contribution from the electroweak sector. Such effective Lagrangian violates the CP symmetry in the strong interaction sector leading to an electric moment dipole to the neutron proportional to $\bar{\theta}$ \cite{Baluni:1978rf,Crewther:1979pi}. Albeit, current measurements of the neutron electric dipole moment indicates that $\bar{\theta}<10^{-10}$ \cite{Baker:2006ts}. This is the strong CP problem, to understand why such parameter should be so small. The Peccei-Quinn solution to this problem is to extend the Standard Model in order have a global chiral anomalous U(1)$_{PQ}$ symmetry, which is spontaneously broken at an energy scale $f_A$, allowing effectively to drive $\bar{\theta}\rightarrow 0$ \cite{Peccei:1977hh,Peccei:1977ur}. The axion is the pseudo-Nambu-Goldstone boson of the U(1)$_{PQ}$ broken symmetry \cite{Weinberg:1977ma,Wilczek:1977pj}, having a mass $m_a= 5.70(7)\,{\rm meV}\left(\frac{10^9\,{\rm GeV}}{f_a}\right)$~\cite{diCortona:2015ldu,Borsanyi:2016ksw}. For a class models \cite{Kim:1979if,Shifman:1979if,Dine:1981rt,Zhitnitsky:1980tq}, the axions have both the mass and the couplings to the Standard Model fields suppressed by the scale $f_A$ and axions can comprise the required relic abundance for being a light cold dark matter candidate  \cite{Preskill:1982cy,Abbott:1982af,Dine:1982ah}. The mass of the axion dark matter mass depends on the assumed cosmological scenario for the U(1)$_{PQ}$ symmetry breaking \cite{Borsanyi:2016ksw}. It has been argued that if the breaking occurs before or during the inflationary period and not restored afterwards the axion dark matter mass range is $10^{-12}\, {\rm{eV}}\leq m_A \leq 10^{-2}\, \rm{eV}$; and if the final symmetry breaking occurs after the inflationary period there is a lower bound $ m_A \geq 28(2)\, \mu\rm{eV}$   \cite{Borsanyi:2016ksw}. Axions have been searched in many experiments and are also the main target of several experimental proposals \cite{vanBibber:2001ud,Bertone:2018xtm,Graham:2015ouw,Ringwald:2016yge}. Different  theoretical studies in the Brazilian community of high energy physics, astrophysics and cosmology have been dedicated to axions and axion-like particles, which are not related to the strong CP problem but have some similar properties with axions, that are in the scope of the actual and near-future experiments \cite{Dias:2014osa,Carvajal:2015dxa,Alves:2016bib,Dias:2018ddy,Carneiro:2018pnc,Alves:2019xpc}.

    \item {\bf Non-Thermal Dark Matter}
    
    Non-thermal dark matter is a broad class of dark matter candidates, which are not axion-like particles but are also produced non-thermally. If one enters in the world of non-thermal processes, the range for dark matter masses and interaction strength is quite vast, conversely to what occurs in the WIMP paradigm. Nevertheless, it is important to keep open-minded because non-thermal dark matter candidates can still fulfill the aforementioned requirements and therefore constitute viable dark matter candidates. A freeze-in production mechanism  \cite{Bernal:2017kxu} and an early matter-dominated universe are some of the avenues that could be explored. In Fig.~\ref{fig:proudBrazilians}, we outline the expected mass range for dark matter candidates that are embedded in our classification of non-thermal dark matter.  
    
\end{itemize}

Unveiling the nature of dark matter is one of the foremost open questions in fundamental science today. This task requires effort from multiple communities passing through nuclear physics, condensed matter, astrophysics, particle physics and cosmology, involving both experimentalists as well as theorists. Over the past decades, we have focused our effort on dark matter particles that experienced a thermal history which have mass in the GeV-TeV range with electroweak interactions with Standard Model Particles. These assumptions were important to guide experimental efforts but with null results thus far motivate us to go beyond the vanilla scenarios. New dark matter candidates have emerged with masses much below the GeV scale and larger than the TeV scale as a result. Dark matter candidates remain well-motivated and we plan to devote time probing them using data from ongoing and upcoming experiments. We intend to explore new avenues that require low energy threshold detectors and maybe new techniques to join the international community agenda. That said, three fronts will be targeted as we outline below:

\section{Direct Detection}

Direct detection refers to the observation of nuclear recoils at low energy experiments located at underground labs. By measurement the energy recoil and the shape of the scattering rate of the events one may reconstruct the dark matter-nucleon (or dark matter-electron) scattering cross-section as well as the dark matter mass. Using different targets and readout techniques several experiments have been proposed. In what follows, we describe the ones that feature Brazilian involvement.

\begin{itemize}

\item  \textbf{DARKSIDE}

The DARKSIDE experiment \cite{Aalseth:2017fik} has Liquid Argon (LAr) as its target and efficiently searches for dark matter using scintillation and ionization readout techniques \cite{Agnes:2018fwg}. After a dark matter-Argon scattering, a prompt scintillation light is emitted ($S_1$ signal), and an ionization process occurs at the same time. In the latter, electrons are drifted towards the anode of the time projection chamber finding a gas phase of Argon where they emit electroluminescence light ($S_2$ signal) \cite{Agnes:2018ves}. Taking advantage of the high trigger efficiency of the $S_2$ signal even for low energy recoils \cite{Agnes:2018oej}, one may use the $S_2$ signal only to constrain dark matter-electron scattering and produce competitive bounds in the $20$~MeV-$1$~GeV dark matter mass window. Its current results are based in 50 Kg of LAr, and the collaboration is now commissioning a 20 tonnes LAr detector.

\item  \textbf{DAMIC}

The  DAMIC  (Dark  Matter  in  Charged Coupled Device) experiment that uses Silicon has a very low energy threshold, and it is suited for light dark matter searches \cite{Aguilar-Arevalo:2016ndq}. The CCDs used in DAMIC were from an existing design for the Dark Energy Survey (DES) camera. Exploring the charge resolution of the device DAMIC is capable of detecting low energy events produced by dark matter \cite{Aguilar-Arevalo:2016zop}. The ionization signal produced in the dark matter-electron scattering can be properly related to the dark matter scattering cross-section and mass. In this way, DAMIC has been able to improve previous bounds on hidden-photon dark matter models \cite{Aguilar-Arevalo:2019wdi}.

    \item  \textbf{CONNIE}

The CONNIE experiment is a CCD (Charged Coupled Device) detector located $30$~m from the core of a commercial nuclear reactor and it has a $1$~keV energy threshold \cite{Aguilar-Arevalo:2016qen}. It has collected 3.7 kg-day of exposure to probing coherent neutrino-nucleus elastic scattering. Particles that couple to neutrinos and quarks contribute to neutrino-nucleus coherent scattering. In particular, a light particle might induce sizeable contribution to the now measured neutrino-nucleus coherent scattering. Such light mediators are also rather popular in dark sector studies. Therefore, CONNIE can test dark sectors.  Indeed, CONNIE obtained a world-leading constraint on a vector mediator with mass $m_{Z^\prime}<10$~MeV \cite{Aguilar-Arevalo:2019zme}.

\item  \textbf{SBND}

The SBND (Short-Baseline Neutrino Detector) constitutes an exciting possibility to probe dark sectors. Light mediators that mix with the photon may be produced in neutrino beams by meson decays and then be detected. Similarly to CONNIE, such neutrino detectors will not probe the dark matter particle itself but might provide complementary and important information concerning the viable dark matter interactions within a dark sector. The dark sector signal is similar to the one produced by neutrino coherent scattering and thus neutrinos constitute the main background. Spectral information and triggers on highly off-axis beams may give a handle on the neutrino background \cite{Machado:2019oxb}.

\item \textbf{DUNE}

The Deep Underground Neutrino Experiment (DUNE) comprises a large far detector (40 kton of liquid argon) placed in the SURF underground laboratory (Lead, South Dakota) and a multi-technology near detector just downstream of the beamline, at Fermilab (Batavia, Illinois) \cite{Abi:2018dnh}. DUNE aims to shed light on critical open questions in neutrino oscillation physics such as CP violation in the lepton sector and the neutrino mass ordering. DUNE may also probe dark matter models that involve neutrino interactions. For instance, one could use DUNE to search for dark matter observing neutrinos from the Sun \cite{Rott:2019stu}, light dark matter \cite{DeRomeri:2019kic} exploring the ability of the DUNE-PRISM near detector of being shifted to different off-axis regions of the neutrino beam. Moreover, missing energy measurements can be done to search for sub-GeV dark matter \cite{Kelly:2019wow}, among other possibilities \cite{Grossman:2017qzw}. Summarizing, due to the size, low-level background of the far detector, cutting edge technology, and concept of the near detector, DUNE has a vast potential to explore dark sectors.

\end{itemize}

\section{Indirect Detection}

Indirect detection of dark matter refers to the observation of stable particles fluxes such as electrons, protons, neutrinos and gamma-rays produced by dark matter annihilation or decay in dense astrophysical environments such as Dwarf Spheroidal Galaxies and the galactic center. Taking the flux of gamma-rays produced from dark matter annihilation as an example, it is known that this flux is proportional to the dark matter annihilation cross-section, energy spectrum (photons produced per annihilation) and the dark matter density along the line of sight, and inversely proportional the dark matter mass. Therefore, if we somehow know the energy spectrum and dark matter density in the study we can correlate the observed flux to the dark matter annihilation cross-section and mass. There are two facilities with Brazilian involvement that are dedicated to probing such dark matter interactions namely CTA and AMS-02. There is also a projected telescope with the goal of measuring Baryon Acoustic Oscillations (BAO) which can provide complementary information concerning dark sectors as we outline below:

\begin{itemize}
    \item \textbf{CTA}

The Cherenkov Telescope Array (CTA) will be the main observatory for high energy gamma-rays in the foreseeable future. The scientific agenda of CTA is quite broad \cite{Acharya:2017ttl}, but one of its major goals is the discovery of dark matter. With instrumental improvement, wider field of view, large energy window covering gamma-rays from $20$~GeV to $300$~TeV, CTA will surpass its predecessors in many ways, and it will be sensitive to dark matter interactions more than one order of magnitude better than current instruments. We highlight that the dark matter annihilation cross-section ($\sigma v = 3\times 10^{-26}\, {\rm cm}^3{\rm s}^{-1}$) within reach of CTA is natural in several dark matter models \cite{Balazs:2017hxh}.  The observatory will operate arrays on sites located in both North and South hemispheres to provide full sky coverage and maximize the discovery potential for the rarest phenomena. In summary, CTA has the potential to discover dark matter via the observation of high energy gamma-rays \cite{Acharya:2017ttl} and for this reason, CTA is one of the flagship experiments. 



\item \textbf{AMS-02}

The Alpha Magnetic Spectrometer is an experiment assembled at International Space Station \cite{Aguilar:2013qda}. It measures antimatter in cosmic rays with unprecedented precision helping understanding the astrophysical processes behind the production of cosmic-rays as well as dark matter. The detector uses several modules that help discriminate gamma-rays from cosmic-rays. In the past years, AMS has observed convincing evidence for a rise in the cosmic-ray positron-electron ratio at energies larger than $10$~GeV, which exceeds significantly the one predicted by secondary positron production  \cite{Aguilar:2019owu}. This signal triggered several dark matter interpretations that have implications
in gamma-ray observations. The continuation of the AMS mission will be important to unveil the origin of the puzzling signal.

\item \textbf{BINGO}

The aforementioned experiments can probe the dark matter interactions and the dark matter density profile, but there are telescopes sensitive to dark sectors in a more general sense. They do not probe the dark matter interactions but they can constrain cosmological parameters which are related to the overall dark matter density. BINGO (BAO from Integrated Neutral Gas Observations) is  a  project of a Radio  Telescope  designed to measure the  neutral  Hydrogen  21  cm  line  via intensity  mapping \cite{Battye:2012tg,Wuensche:2019cdv}. In order words, its primary goal is to measure BAO  at  radio frequencies.  BAO is one of the main evidence for dark matter and therefore its observation has direct implications to dark sectors \cite{Wang:2016lxa}. 

\end{itemize}

\section{Colliders}

The Large Hadron Collider (LHC) collides proton beams at very high energy to probe the fundamental interactions in Nature. It discovered the Higgs boson as predicted by the Standard Model in 2012. Many particles might be produced during these collisions, including dark matter. Dark matter particles interact weakly with matter and for this reason, leave no trace at the detectors. Therefore, one can use energy-momentum conservation to infer the presence of dark matter in the final state of proton-proton collisions. In other words, they represent missing energy. Generally speaking, missing energy searches are dark matter searches. It is important to highlight that LHC cannot affirm that the missing energy observed is attributed to dark matter particles as any neutral long-lived particles would mimic the same signal. Anyhow, colliders constitute important probes for dark matter as they are not subject to large astrophysical and nuclear uncertainties as direct and indirect detection searches and indeed give rise to important bounds on the parameter space of several dark matter models. We describe below the colliders featuring Brazilian involvement:

\begin{itemize}

\item \textbf{ATLAS}

ATLAS (A Toroidal LHC ApparatuS) is a particle detector experiment at the Large Hadron Collider (LHC) located at CERN (The European Organization for Nuclear Research). ATLAS searches for dark sectors by probing either the mediators that connect dark matter particles with the Standard Model ones or the dark matter particles itself. That said, the mediator could have sizeable couplings to leptons and quarks. Therefore, one could probe this mediator searching for dijet and dilepton resonances. Such mediator might have a large branching ratio into dark matter affecting the overall bound on the mediator mass from dijet and dilepton searches \cite{Aaboud:2017yvp,Aaboud:2017buh}. Thus, one can explore the connection between the dark matter, mediator and Standard Model particles in this way. The other possible way consists of searching for the pair-production of dark matter in association with a visible object such as a jet, photon, etc, typically from initial state radiation . These searches are known as mono-X searches, but they might contain more than one visible object \cite{Aaboud:2017phn}. Such a visible object is necessary because the production of dark matter alone is invisible to the detector but signal containing dark matter and visible objects will appear as an imbalance in the transverse momentum of the event ($p_\textrm{T}^\textrm{miss}$). Therefore, dark matter signals are characterized by large missing energy \cite{Alpigiani:2017juv}. In a similar vein searches for long-lived particles that belong to a dark sector can also be performed \cite{Aaboud:2018jbr}.



\item \textbf{CMS}

CMS (Compact Muon Solenoid) is a general-purpose particle detector at the LHC. 
Similarly to ATLAS, it has a broad physics program that covers from Standard Model precision measurements to searches for new physics, including dark matter. 
Despite having the same scientific goals as the ATLAS experiment, the CMS collaboration adopted different technical solutions for the design, construction and operation of its experiment. 
Thus, CMS is able to provide independent and complementary new physics searches. 

The mainline dark matter searches in CMS extract the signal by fitting the missing transverse momentum distribution of events passing the mono-X criteria, in a similar vein as the ATLAS experiment~\cite{Sirunyan:2017jix,Sirunyan:2017qfc,Sirunyan:2018gdw,Sirunyan:2019gfm}. 
In particular, the missing transverse momentum threshold for the online data acquisition is set at about $100$~GeV. 
The two main SM backgrounds, $Z(\nu\nu)$ and $W(\ell\nu)$ processes, are estimated using a set of control regions enriched in leptons and photons. 
The modelling of the $p_\textrm{T}^\textrm{miss}$ distribution is dominated by systematic uncertainties for low values (close to the $100$~GeV threshold), whilst statistical uncertainties are the main factor for high value (close to the TeV scale)~\cite{Vartak:2017yfz}.

The CMS collaboration interprets its results in terms of simplified dark matter models and compares the findings with the results stemming from the direct detection experiments.
CMS also searches for evidence for dark matter using the long-lived particle signatures~\cite{Sirunyan:2018njd,Sirunyan:2019gut}, and recasts their dijet and dilepton resonance searches as dark matter mediator searches~\cite{Sirunyan:2017nvi,Sirunyan:2018wcm,Sirunyan:2019vgj}. 


\item \textbf{LHCb}

The LHCb (Large Hadron Collider beauty) experiment is one of the detectors at the LHC. LHCb is dedicated to b-physics but has searched for dark matter candidates probing different models as well. In general, the strategy is to search for long-lived particles or dimuon resonances.
In general, the LHCb searches focus on low mass with short lifetimes due to the lower thresholds and integrated luminosity compared to ATLAS and CMS.

The direct production of long-lived particle searches are based on data selected by triggers on displaced vertices with a transverse distance greater than $4$~ mm with four or more tracks or with muons~\cite{LHCb-PAPER-2016-065,LHCb-PAPER-2016-047}
. One of the main background sources is the hadronic interactions with the detector that can mimic long-lived particle decays.
The benchmark model used by these searches is a scalar particle decaying to two neutral long-lived particles (dark or “valley” pions), which corresponds to the Higgs simplified model with hadronic decay modes.

The direct production of both promptly decaying and long-lived dark photons are studied at LHCb using two low transverse momentum muons~\cite{LHCb-PAPER-2017-038,LHCb-PAPER-2019-031}. The dark photons do not tend to be highly boosted in the transverse direction, therefore the limits are competitive even at higher mass. 
The searches for displaced leptons in rare B meson decays ~\cite{LHCb-PAPER-2015-036,LHCb-PAPER-2016-052} are performed by identifying the $B^{\pm}$ decay vertex and the dimuon invariant mass is scanned for excesses.

In LHC Run 3, LHCb experiment plans to take data at an instantaneous luminosity five times larger than the previous Run. The hardware trigger will be removed and a full event reconstruction will be performed at the bunch crossing rate of the LHC with a flexible software-based trigger. Projections of the LHCb analyses sensitivity are promising and demonstrates the great importance of the LHCb detector to fill in the gaps in the search program of ATLAS and CMS~\cite{Alimena:2019zri,CidVidal:2018eel}

\item \textbf{Future Colliders}

Several Brazilian groups are involved in the  HL-LHC (High-Luminosity-LHC) and HE-LHC (High-Energy-LHC) upgrade which will significantly improve our potential to discover physics beyond the Standard Model. The HL-LHC upgrade is defined as the phase under which an integrated luminosity of $\mathcal{L}=3$~ab$^{-1}$  will be achieved with a center-of-mass energy of $14$~TeV. A possible and important upgrade is the HE-LHC where the luminosity, as well as the center-of-mass energy, will be increased to $\mathcal{L}=15$~ab$^{-1}$ and  $27$~TeV, respectively \cite{CidVidal:2018eel}. In the process of the LHC upgrade another collider, the Large Hadron electron Collider (LHeC) may be built. It is capable of probing new regions of the parameter space taking advantage of the electron polarization \cite{Lindner:2016lxq,DOnofrio:2019dcp}. There are other future collider proposals that also have the potential to significantly improve our understanding about the laws of nature and probe dark matter models \cite{No:2019gvl,deBlas:2018mhx,Kalinowski:2018kdn} including searches for dark matter in double Higgs production which is a major goal of future colliders~\cite{Banerjee:2016nzb,Alves:2019emf}.

\end{itemize}

\section{Theoretical Efforts}

The Brazilian community is involved in dark matter physics both on the experimental and theoretical perspectives. Despite the wealth of evidence on the existence of dark matter, we do not know its particle nature and many of its astrophysical properties. There are astrophysical aspects of dark matter which are open to debate, in particular at small astrophysical scales. Among them, the most persistent one is the diversity problem (which is a generalization of the famous cusp-core issue) \cite{Moore:1994yx,Flores:1994gz,deBlok:2009sp,Oman:2015xda,Santos-Santos:2019vrw}. Another one, known for a long time is the regularity issue, as inferred from the mass-discrepancy-acceleration relation (also called radial acceleration relation) \cite{McGaugh:2016leg,Navarro:2016bfs,Ren:2018jpt}. The underlying causes of such problems need to be addressed. Its understanding could promote development towards either dark matter itself (e.g., halo profiles, whether it is warm, self-interacting, ultra-light) or for the understanding of baryonic astrophysical processes (e.g., stellar population evolution, active galactic nucleus). Indeed, the Brazilian community has the know-how to model and infer the dark matter consequences from large samples of galaxies \cite{Rodrigues:2014xka} and more recently using Bayesian inference \cite{Rodrigues:2018duc}. These works are relevant for the discrepancies observed at small scales, covering the diversity/cusp-core problem (halo profiles) and the possible fundamental nature of the radial acceleration relation  \cite{Rodrigues:2014xka,Rodrigues:2017vto}. At larger scales, different methods are under consideration in order to predict the clustering of dark matter at non-linear level in the presence of dynamical dark energy \cite{Casarini:2016ysv}. At linear perturbative level, the assumptions considered include dark matter as a dissipative component due to a viscosity coefficient \cite{Velten:2014xca}, with a velocity dispersion on the evolution of its perturbations \cite{Piattella:2015nda}, or with warmness represented by a reduced relativistic gas \cite{Hipolito-Ricaldi:2017kar,Pordeus-da-Silva:2019bak} 

Considering the dark matter profile, also the importance of the halo asphericity was stressed by \cite{Bernal:2014mmt}. Uncertainties on the overall dark matter density are also explored in the models beyond the $\Lambda$CDM \cite{Pigozzo:2015swa,deAlmeida:2016chs,Marttens:2017njo,Ren:2018jpt,vonMarttens:2018iav}. It has been noticed that one might indeed have deviations from the cold dark matter abundance inferred by Planck in models where dark matter and dark energy are coupled \cite{Davari:2019tni,vonMarttens:2019ixw}

On the particle physics side, several plausible new dark matter models have been proposed going from WIMPs \cite{Mizukoshi:2010ky}, axions \cite{Alves:2016bib,Montero:2017yvy} to non-thermal dark matter \cite{Bhattacharya:2019ucd}, where the latter has an exciting connection with the IceCube data. The neutrino messenger in the indirect search for dark matter is a promising channel. This channel can be used either in the searches for neutrinos from compact objects, such as the Sun \cite{Esmaili:2010wa,Esmaili:2009ks}, or possible contribution to the diffuse high energy neutrino flux observed by IceCube \cite{Esmaili:2013gha,Esmaili:2014rma}. A popular model in the context of non-thermal dark matter is the so-called dark photon which is subject to intensive searches at accelerators and low energy colliders. Alternatively, astrophysical observations have become a well-known tool to obtain empirical constraints on dark photons. Generally speaking, any light particle can play an important
role in stellar energy loss. Such a particle would remove energy from the stellar thermal bath.  If a stellar
matter has a sufficient content of dark matter, an important process to consider is the emission of dark bosons (axions, dark photons etc) from thermal states \cite{Santos:2014xka,Corsico:2012ki}. Thus there is a rich phenomenology to be explored in this context.

Considering the experimental and theoretical aspects raised in this report it is clear that dark matter is under siege by the Brazilian community.


\bibliographystyle{JHEPfixed}
\bibliography{references}
\end{document}